\documentstyle[aps,prl,epsf,floats]{revtex}
\newcommand{\beq}{\begin{equation}}
\newcommand{\eeq}{\end{equation}}
\newcommand{\beqs}{\begin{eqnarray}}
\newcommand{\eeqs}{\end{eqnarray}}

\newcommand{\lsim}{\mathrel{\raisebox{-.6ex}{$\stackrel{\textstyle<}{\sim}$}}}
\newcommand{\gsim}{\mathrel{\raisebox{-.6ex}{$\stackrel{\textstyle>}{\sim}$}}}

\draft

\tighten

\begin{document}

\twocolumn[\hsize\textwidth\columnwidth\hsize\csname
@twocolumnfalse\endcsname

\title{Dynamical Symmetry Breaking of Extended Gauge Symmetries}

\author{Thomas Appelquist$^1$ and Robert Shrock$^2$}
\address{$^1$ Department of Physics, Sloane Laboratory, Yale University,
New Haven, CT 06520 \\
C. N. Yang Institute for Theoretical Physics, State University of New York,
Stony Brook, NY  11794}

\baselineskip 6.0mm

\tighten

\maketitle

\begin{abstract}

We construct asymptotically free gauge theories exhibiting dynamical
breaking of the left-right, strong-electroweak gauge group $G_{LR} = {\rm
SU}(3)_c \times {\rm SU}(2)_L \times {\rm SU}(2)_R \times {\rm U}(1)_{B-L}$,
and its extension to the Pati-Salam gauge group $G_{422}={\rm SU}(4)_{PS}
\times {\rm SU}(2)_L \times {\rm SU}(2)_R$. The models incorporate
technicolor for electroweak breaking, and extended technicolor for the
breaking of $G_{LR}$ and $G_{422}$ and the generation of fermion masses,
including a seesaw mechanism for neutrino masses. These models explain
why $G_{LR}$ and $G_{422}$ break to ${\rm SU}(3)_c \times {\rm SU}(2)_L
\times {\rm U}(1)_Y$, and why this takes place at a scale ($\sim 10^3$ TeV)
which is large compared to the electroweak scale.

\end{abstract}

\pacs{12.60.Cn, 12.60.Nz, 14.60.Pq}

\vskip2.0pc]

The standard model (SM) gauge group $G_{SM} = {\rm SU}(3)_c \times {\rm
SU}(2)_L \times {\rm U}(1)_Y$ has provided a successful description of both
strong and electroweak interactions. Although the standard model itself
predicts zero neutrino masses, its fermion content can be augmented to
accomodate the current evidence for neutrino masses and lepton mixing. But the
origin of the electroweak symmetry breaking (EWSB) is still not understood. It
might occur via the Higgs mechanism, as in the SM. An alternative is dynamical
symmetry breaking (DSB) of the electroweak symmetry, driven by a strongly
coupled, asymptotically-free, vectorial gauge interaction associated with an
unbroken gauge symmetry, denoted generically as technicolor (TC)
\cite{tcrev}-\cite{scalc}.

There has also long been interest in models with gauge groups larger than
$G_{SM}$. One such model has the gauge group \cite{lrs}
\beq
G_{LR} = {\rm SU}(3)_c \times {\rm SU}(2)_L \times {\rm SU}(2)_R \times
{\rm U}(1)_{B-L}
\label{glrs}
\eeq
in which the fermions of each generation transform as $(3,2,1)_{1/3,L}$,
$(3,1,2)_{1/3,R}$, $(1,2,1)_{-1,L}$, and $(1,1,2)_{-1,R}$.  The gauge couplings
are defined via the covariant derivative $D_\mu = \partial_\mu - ig_3 {\bf T}_c
\cdot {\bf A}_{c,\mu} -ig_{2L} {\bf T}_L \cdot {\bf A}_{L,\mu} -ig_{2R} {\bf
T}_R \cdot {\bf A}_{R,\mu} -i(g_{_U}/2)(B-L){\bf U}_\mu$. In this model the
electric charge is given by the elegant relation $Q = T_{3L} + T_{3R} +
(B-L)/2$, where $B$ and $L$ denote baryon and (total) lepton number. $G_{LR}$
would break at a scale $\Lambda_{LR}$ well above the electroweak scale.

The model based on $G_{LR}$ may be further embedded in a model with gauge
group \cite{ps}
\beq
G_{422} = {\rm SU}(4)_{PS} \times {\rm SU}(2)_L \times {\rm SU}(2)_R \ .
\label{g422}
\eeq
This model provides a higher degree of unification since it combines
U(1)$_{B-L}$ and SU(3)$_c$ (in a maximal subgroup) in the Pati-Salam group
SU(4)$_{PS}$ and hence relates $g_{_U}$ and $g_3$.  Denoting the generators
of SU(4)$_{PS}$ as $T_{PS,i}$, $1 \le i \le 15$, with $T_{PS,15} =
(2\sqrt{6})^{-1}{\rm diag}(1,1,1,-3)$ and setting $U_\mu = A_{PS,15,\mu}$,
one has $(B-L)/2 = \sqrt{2/3}T_{PS,15}$, and hence $\left (g_{_U} / g_{PS}
\right )^2 = 3/2 \quad {\rm at} \ \ \Lambda_{PS}$, where $\Lambda_{PS}$ is
the breaking scale of the $G_{422}$ group. This model also has the appeal
that it quantizes electric charge, since $Q =
T_{3L}+T_{3R}+\sqrt{2/3}T_{PS,15}=T_{3L}+T_{3R} +(1/6){\rm diag}(1,1,1,-3)$.

The conventional approach to the gauge symmetry breaking of these models
employs elementary Higgs fields and arranges for a hierarchy of breaking scales
by making the vacuum expectation values (vev's) of the Higgs that break
$G_{LR}$ or $G_{422}$ to $G_{SM}$ much larger than the Higgs vev's that break
${\rm SU}(2)_L \times {\rm U}(1)_Y \to {\rm U}(1)_{em}$ \cite{lrs,mpsusy}.
This hierarchy is necessitated by the experimental lower limits on the masses
of a possible $W_R$ or $Z^\prime$ \cite{data}.  An interesting question is
whether one can construct asymptotically free gauge theories containing the
group $G_{LR}$ and/or $G_{422}$ that exhibit dynamical breaking of all the
gauge symmetries other than $SU(3)_c$ and $U(1)_{em}$, that naturally explain
the hierarchy of breaking scales, and that yield requisite light neutrino
masses. In this letter, we present such models.

Technicolor itself cannot provide a mechanism for all the breaking, because
it is too weak at the the scale $\Lambda_{LR}$ or $\Lambda_{PS}$ and because
the technifermion condensate $\langle \bar F F \rangle = \langle \bar F_L
F_R \rangle + \langle \bar F_R F_L \rangle$ would break both ${\rm SU}(2)_L$
and ${\rm SU}(2)_R$ at the same scale (to the diagonal (vector) group
SU(2)$_V$). Of course, to explain quark and lepton mass generation and
incorporate the three families, technicolor has to be enlarged to an
extended technicolor (ETC) theory \cite{etc}. Our models are ETC-type
theories, with the breaking of $G_{LR}$ and $G_{422}$ to $G_{SM}$ being
driven by the same interactions that break the ETC group and generate quark
and lepton masses.

Taking the technicolor gauge group to be SU($N_{TC})$, the technifermions
comprise an additional family, viz., $Q_L = {U \choose D}_L$, $L_L = {N
\choose E}_L$, $U_R$, $D_R$, $N_R$, $E_R$ transforming according to the
fundamental representation of SU($N_{TC}$) and the usual representations of
$G_{SM}$ (where color and TC indices are suppressed). Vacuum alignment
considerations yield the desired color- and charge-conserving TC condensates
\cite{valign}. To satisfy constraints from flavor-changing neutral-current
processes, the ETC vector bosons that can mediate generation-changing
transitions must have large masses. We envision that these arise from
self-breaking of an ETC gauge symmetry, which requires that ETC be a
strongly coupled, chiral gauge theory. The self-breaking occurs in stages,
for example at the three stages $\Lambda_1 \sim 10^3$ TeV, $\Lambda_2 \sim
50$ TeV, and $\Lambda_3 \sim 3$ TeV, corresponding to the $3$ standard-model
fermion generations. Hence $N_{ETC}=N_{TC}+3$.

A particularly attractive choice for the technicolor group, used in the models
studied here, is ${\rm SU}(2)_{TC}$, which thus entails $N_{ETC}=5$.  With $N_f
= 8$ vectorially coupled technifermions in the fundamental representation,
studies suggest that this SU(2)$_{TC}$ theory could have an (approximate)
infrared fixed point (IRFP) in the confining phase with spontaneous chiral
symmetry breaking \cite{vals,gap}. This approximate IRFP produces a slowly
running (``walking'') TC gauge coupling, which can yield realistically large
quark and charged lepton masses \cite{wtc}.  The choice $N_{TC}=2$ and the
walking can strongly reduce TC contributions to the $S$ parameter
\cite{scalc,nutev}. Further ingredients may be needed to account for the 
top-quark mass. 

In Ref. \cite{nt}, we studied the generation of neutrino masses in an ETC model
of this sort and showed that light neutrino masses and lepton mixing can be
produced via a seesaw without any superheavy mass scales. Here we extend this
model to the groups $G_{LR}$ and $G_{422}$.

We recall that $\Lambda_{TC}$ is determined by using the relation $m_W^2 =
(g^2/4)(N_c f_Q^2 + f_L^2) \simeq (g^2/4)(N_c+1)f_F^2$, where for our purposes
we take $f_L \simeq f_Q \equiv f_F$.  This gives $f_F \simeq 130$ GeV.  In
QCD, $f_\pi = 93$ MeV and $\Lambda_{QCD} \sim 170$ MeV, so that
$\Lambda_{QCD}/f_\pi \sim 2$; using this as a guide to technicolor, we infer
$\Lambda_{TC} \sim 260$ GeV.  The induced fermion masses in the $i$'th
generation are given by $m_{f_i} \sim g_{_{ETC}}^2 \eta_i
N_{TC}\Lambda_{TC}^3/(4\pi^2 M_i^2)$, where $M_i \sim g_{_{ETC}}\Lambda_i$ is
the mass of the ETC gauge bosons that gain mass at scale $\Lambda_i$ and
$g_{_{ETC}}$ is the running ETC gauge coupling evaluated at this scale.  The
quantity $\eta_i$ is a possible enhancement factor incorporating walking, for 
which $\eta_i \sim \Lambda_i/f_F$ \cite{wtc,eta}.

We first consider the standard-model extension based on $G_{LR}$. Our model
for the DSB utilizes the gauge group
\beq G = {\rm SU}(5)_{ETC} \times {\rm SU}(2)_{HC} \times G_{LR} \label{g}
\eeq
where HC denotes hypercolor, a second strong gauge interaction which,
together with ETC, triggers the requisite sequential breaking pattern.
The fermion content of this model is listed below; the numbers
indicate the representations under ${\rm SU}(5)_{ETC} \times {\rm
SU}(2)_{HC} \times {\rm SU}(3)_c \times {\rm SU}(2)_L \times {\rm SU}(2)_R$
and the subscript gives $B-L$:
\beqs
& & (5,1,3,2,1)_{1/3,L} \ , \quad (5,1,3,1,2)_{1/3,R} \ , \cr\cr
& & (5,1,1,2,1)_{-1,L}  \ , \quad (5,1,1,1,2)_{-1,R} \ , \cr\cr
& & (\bar 5,1,1,1,1)_{0,R} \ , \quad (\overline{10},1,1,1,1)_{0,R} \ , \quad
(10,2,1,1,1)_{0,R} \ .
\label{lrfermions}
\eeqs
Thus the fermions include a vectorlike set of quarks and techniquarks in the
representations $(5,1,3,2,1)_{1/3,L}$, $(5,1,3,1,2)_{1/3,R}$ and leptons and
technileptons in $(5,1,1,2,1)_{-1,L}$, $(5,1,1,1,2)_{-1,R}$, together with a
set of $G_{LR}$-singlet fermions in $(\bar 5,1,1,1,1)_{0,R}$,
$(\overline{10},1,1,1,1)_{0,R}$, and $(10,2,1,1,1)_{0,R}$ \cite{rh}. The
leptons and technileptons are denoted $L^{i,p}_\chi$, where $\chi=L,R$, $1
\le i \le 5$, and $p=1,2$. The $G_{LR}$-singlets are denoted respectively
${\cal N}_{i,R}$, $\psi_{ij,R}$, and $\zeta^{ij,\alpha}_R$, where $1 \le i,j
\le 5$ are ETC indices and $\alpha,\beta$ are SU(2)$_{HC}$ indices.  The 
models with $G_{LR}$ and $G_{422}$ share several features with the ETC model
in \cite{at94}. 

The SU(5)$_{ETC}$ theory is an anomaly-free, chiral gauge theory and, like
the ETC and HC theories, is asymptotically free.  There are no bilinear
fermion operators invariant under $G$, and hence there are no bare fermion
mass terms.  The SU(2)$_{HC}$ and SU(2)$_{TC}$ subsectors of SU(5)$_{TC}$
are vectorial.

To analyze the stages of symmetry breaking, we identify plausible preferred
condensation channels using a generalized-most-attractive-channel (GMAC)
approach that takes account of one or more strong gauge interactions at each
breaking scale, as well as the energy cost involved in producing gauge boson
masses when gauge symmetries are broken. In this framework, an approximate
measure of the attractiveness of a channel $R_1 \times R_2 \to R_{cond.}$ is
$\Delta C_2 = C_2(R_1)+C_2(R_2)-C_2(R_{cond.})$, where $R_j$ denotes the
representation under a relevant gauge interaction and $C_2(R)$ is the
quadratic Casimir.

As the energy decreases from some high value, the SU(5)$_{ETC}$ and
SU(2)$_{HC}$ couplings increase. We envision that at $E \sim \Lambda_{LR}
\gsim 10^3$ TeV, $\alpha_{_{ETC}}$ is sufficiently strong \cite{gap} to produce
condensation in the channel
\beq
(5,1,1,1,2)_{-1,R} \times (\bar 5,1,1,1,1)_{0,R} \to (1,1,1,1,2)_{-1}
\label{55barchannel}
\eeq
with $\Delta C_2 = 24/5$, breaking $G_{LR}$ to ${\rm SU}(3)_c \times {\rm
SU}(2)_L \times {\rm U}(1)_Y$.  The associated condensate is
$\langle L^{i,p \ T}_R C {\cal N}_{i,R} \rangle$, where $1 \le i \le 5$ is
an SU(5)$_{ETC}$ index and $p \in \{1,2\}$ is an SU(2)$_R$ index.  With no
loss of generality, we use the initial SU(2)$_R$ invariance to rotate the
condensate to the $p=1$ component, $L^{i,p=1}_R \equiv n^i_R$, which is
electrically neutral and has weak hypercharge $Y=0$; the condensate is thus
$\langle n^{i \ T}_R C {\cal N}_{i,R} \rangle$ so that the $n^i_R$ and
${\cal N}_{i,R}$ gain dynamical masses $\sim \Lambda_{LR}$.

There exists a more attractive channel than (\ref{55barchannel}) in a simple
MAC analysis: $(\overline{10},1,1,1,)_{0,R} \times (10,2,1,1,)_{0,R} \to
(1,2,1,1,)_0$, with $\Delta C_2=36/5$. But with the coupling $g_{_{HC}}$
also large at $\Lambda_{LR}$, a sizeable energy price would be incurred in 
this channel to generate the vector boson masses associated with the
breaking of the SU(2)$_{HC}$. We assume here that this price is higher than
the energy advantage due to the greater attractiveness of the channel
$(\overline{10},1,1,1,)_{0,R} \times (10,2,1,1,)_{0,R} \to (1,2,1,1,)_0$
\cite{GMAC}.

The condensation (\ref{55barchannel}) generates masses
\beq m_{W_R} = \frac{g_{2R}}{2}\Lambda_{LR} \quad\quad m_{Z^\prime} =
\frac{g_{2u}}{2}\Lambda_{LR} \ , \label{mwrmzprime} \eeq
where $g_{2u} \equiv \sqrt{g_{2R}^2 + g_{_U}^2}$, for the $W^\pm_{R,\mu} =
A^\pm_{R,\mu}$ gauge bosons and the linear combination
\beq Z^\prime_\mu = \frac{g_{2R}A_{3,R,\mu}-g_{_U} U_\mu}{g_{2u}} .
\label{zprime} \eeq
This leaves the orthogonal combination
\beq
B_\mu=\frac{g_{U} A_{3,R,\mu}+g_{2R} U_\mu}{g_{2u}}
\label{b}
\eeq
as the weak hypercharge U(1)$_Y$ gauge boson, which is massless at this stage.
The hypercharge coupling is then
\beq
g^\prime = \frac{g_{2R}g_{_U}}{g_{2u}} \ .
\label{gprime}
\eeq
so that, with $e^{-2}=g_{2L}^{-2}+(g^\prime)^{-2}=
g_{2L}^{-2}+g_{2R}^{-2}+g_{_U}^{-2}$, the weak mixing angle is given by
\beq
\sin^2 \theta_W = \biggl [ 1 + \Bigl (\frac{g_{2L}}{g_{2R}} \Bigr )^2
+ \Bigl (\frac{g_{2L}}{g_{_U}} \Bigr )^2 \biggr ]^{-1}
\label{swsq}
\eeq
at the scale $\Lambda_{LR}$. The experimental value of $\sin^2 \theta_W$ at
$M_Z$ can be accommodated naturally, for example with all couplings in
(\ref{swsq}) of the same order ( even with $g_{2R} = g_{2L}$) and with
modest RG running from $\Lambda_{LR}$ to $M_Z$.

For $E < \Lambda_{LR}$, the fermion content of the effective theory is
\beqs
& & (5,1,3,2)_{1/3,L} \ , \quad (5,1,3,1)_{4/3,R} \ , \quad
(5,1,3,1)_{-2/3,R}\cr\cr
& & (5,1,1,2)_{-1,L}  \ , \quad (5,1,1,1)_{-2,R} \ , \cr\cr
& & (\overline{10},1,1,1)_{0,R} \ , \quad (10,2,1,1)_{0,R} \ ,
\label{lrfermionsbelowlamlr}
\eeqs
where the entries refer to ${\rm SU}(5)_{ETC} \times {\rm SU}(2)_{HC} \times
{\rm SU}(3)_c \times {\rm SU}(2)_L$ and $Y$ is a subscript. This is
precisely the gauge group and fermion content of the ETC model that we
analyzed in Ref. \cite{nt} with a focus on the formation of neutrino masses.
We therefore summarize the subsequent stages of breaking only briefly,
drawing on results of \cite{nt}.

At a value $E \sim \Lambda_1 \sim 10^3$ TeV comparable to $\Lambda_{LR}$, a
GMAC analysis suggests that there is condensation in the channel
\beq
(\overline{10},1,1,1)_{0,R} \times (\overline{10},1,1,1)_{0,R} \to
(5,1,1,1)_0 \ .
\label{10b10b5channel}
\eeq
Thus, ${\rm SU}(5)_{ETC}$ self-breaks to ${\rm SU}(4)_{ETC}$, producing
masses $\sim g_{_{ETC}}\Lambda_1$ for the nine gauge bosons in the coset
SU(5)$_{ETC}$/SU(4)$_{ETC}$.  As at $\Lambda_{LR}$, we assume that a GMAC
analysis favors this channel over the $10 \times \overline{10}$ channel in
which $SU(2)_{HC}$-breaking gauge boson masses $\sim g_{_{HC}}\Lambda_1$
would have to be formed. Although the latter channel is more attractive, a
very large energy price would have to be paid for the associated vector
boson mass generation for sufficiently large $\alpha_{_{HC}}
> \alpha_{_{ETC}}$. Also, although (\ref{10b10b5channel}) has the same
$\Delta C_2$-value ($ =24/5$) as (\ref{55barchannel}), it is plausible that
$\Lambda_1 \lsim \Lambda_{LR}$, since an energy price ($\sim
g_{_{ETC}}\Lambda_1$) is incurred by the breaking of $SU(5)_{ETC}$.

The SU(5)$_{ETC} \to$ SU(4)$_{ETC}$ breaking entails the separation of the
first generation of quarks and leptons from the components of SU(5)$_{ETC}$
fermion fields with indices $2 \le i \le 5$. The further ETC gauge symmetry
breaking occurs in stages, leading eventually to the SU(2)$_{TC}$ subgroup
of the original SU(5)$_{ETC}$ group.  We have identified two plausible
sequences for this breaking \cite{at94,nt}. Both sequences yield a strongly
coupled SU(2)$_{TC}$ gauge interaction that produces a TC condensate,
breaking ${\rm SU}(2)_L \times {\rm U}(1)_Y \to {\rm U}(1)_{em}$ \cite{alsv}.

Dirac mass terms for the neutrinos are formed dynamically, involving the
left-handed neutrinos in the $(5,1,1,2,1)_{-1,L}$, but not their respective
right-handed counterparts in the $(5,1,1,1,2)_{-1,R}$. Instead, the
right-handed partners emerge from the $(\overline{10},1,1,1,1)_{0,R}$ (as
$\psi_{1j,R}$, $j=2,3$). Thus there are only two right-handed neutrinos.  In
a model in which $L$ is not gauged, it is a convention how one assigns the
lepton number $L$ to the SM-singlet fields. Here, $L=0$ for the fields that
are singlets under $G_{LR}$ or $G_{422}$, since they are singlets under
U(1)$_{B-L}$ and have $B=0$. Hence, the neutrino Dirac mass terms violate
$L$ by 1 unit. There are also larger, Majorana masses generated for the
$\psi_{ij,R}$ fields themselves; the seesaw mechanism then leads to
left-handed $\Delta L=2$ Majorana neutrino bilinears \cite{lcon}.

We next consider the extension of the standard model gauge group to
$G_{422}$. In this case, our full model is based on the gauge group $G={\rm
SU}(5)_{ETC} \times {\rm SU}(2)_{HC} \times G_{422}$ with fermion content
\beqs
& &  (5,1,4,2,1)_L \ , \quad (5,1,4,1,2)_R \ , \cr\cr
& & (\bar 5,1,1,1,1)_R \ , \quad (\overline{10},1,1,1,1)_R \ , \quad
(10,2,1,1,1)_R \ .
\label{422fermions}
\eeqs
Again, as $E$ decreases from high values, the SU(5)$_{ETC}$ and SU(2)$_{HC}$
couplings increase. At a scale $\Lambda_{PS}$, the SU(5)$_{ETC}$ coupling
will be large enough to produce condensation in the channel
\beq
(5,1,4,1,2)_R \times (\bar 5,1,1,1,1)_R \to (1,1,4,1,2) \ .
\label{55barchannelps}
\eeq
This breaks ${\rm SU}(4)_{PS} \times {\rm SU}(2)_R$ directly to ${\rm
SU}(3)_c \times {\rm U}(1)_Y$.  The value $\Lambda_{PS} \sim 10^3$ TeV
satisfies phenomenological constraints, e.g. from the upper limit on
$BR(K_L \to \mu^\pm e^\mp)$. The associated condensate is again $\langle
n^{i \ T}_R C {\cal N}_{i,R} \rangle$, and the $n^i_R$ and ${\cal N}_{i,R}$
gain masses $\sim \Lambda_{PS}$.  The results
(\ref{mwrmzprime})-(\ref{swsq}) apply with the condition $\left (g_{_U} /
g_{PS} \right )^2 = 3/2 \quad {\rm at} \ \ \Lambda_{PS}$.

Further breaking at lower scales proceeds as in the $G_{LR}$ model and as
described in Ref. \cite{nt}. Dirac mass terms for the neutrinos are formed
from the $(5,1,4,2,1)_L$ and the $(\overline{10},1,1,1,1)_R$, leading to the 
same type of seesaw as in \cite{nt} and the $G_{LR}$ model.

The experimental value of $\sin^2 \theta_W$ can again be accommodated by
(\ref{swsq}), although this now necessarily requires $g_{2R} < g_{2L}$ at
$\Lambda_{PS}$. To see this, we evolve the SM gauge couplings from $\mu=m_Z$ to
the EWSB scale $\Lambda_{EW}=2^{-3/4}G_F^{-1/2}=174$ GeV and then from
$\Lambda_{EW}$ up to $\Lambda_{PS}$ using $d\alpha_j/dt = -b_0
\alpha_j^2/(2\pi) + O(\alpha_j^3) + ...$ where $t=\ln \mu$, $\alpha_1 \equiv
(g^\prime)^2/(4\pi)$, and $...$ denotes theoretical uncertainties associated
with mass thresholds.  In the interval $\Lambda_{EW} \le \mu \le \Lambda_{PS}$
we include the contributions from the $t$ quark and relevant technifermions, so
that $b_0^{(3)} = 13/3$, $b_0^{(2)} = 2/3$, and $b_0^{(1)} = -10$.  The initial
values at $m_Z$ are $\alpha_3(m_Z)= 0.118$, $\alpha_{em}(m_Z)^{-1}=129$, and
$(\sin^2\theta_W)_{\overline{MS}}(m_Z) = 0.231$ \cite{pdg,nutev}.  With
$\Lambda_{PS}=10^6$ GeV and the calculated values $\alpha_3=0.064$,
$\alpha_{2L}=0.032$, $\alpha_1=0.012$ at $\Lambda_{PS}$, we find
$\alpha_{2R}(\Lambda_{PS}) \simeq 0.013$ so that $g_{2R}/g_{2L} \simeq 0.64$ at
this scale.

It may be possible to allow $g_{2R} = g_{2L}$ at $\Lambda_{PS}$, and still
match $(\sin^2 \theta_W)_{exp.}$, by further expanding the (4D) gauge theory to
one with, e.g., ${\rm SU}(4)_{PS} \times {\rm SU}(2)^4$ as in \cite{su424} but
with DSB; we are currently studying this \cite{chp}.

To summarize, we have constructed asymptotically free models with dynamical
symmetry breaking of the extended gauge groups $G_{LR}$ and $G_{422}$. These
models involve higher unification, and $G_{422}$ has the appeal of
quantizing electric charge.  Our models naturally explain why (i) $G_{LR}$
and $G_{422}$ break to $G_{SM}$ and (ii) this breaking occurs at the scales
$\Lambda_{LR}, \ \Lambda_{PS} >> m_{W,Z}$.  The models incorporate
technicolor for electroweak symmetry breaking, and extended technicolor for
fermion mass generation including a seesaw mechanism for the generation of
realistic neutrino masses.

A different approach appears to be needed to construct a theory with
dynamical breaking of the grand unified groups $G_{GUT}={\rm SU}(5)$ or
SO(10) because, among other things, if the ETC group commuted with
$G_{GUT}$, then, with the standard fermion assignments in these GUT groups,
the quarks and charged leptons would not transform in a vectorial manner
under $G_{ETC}$, so that the usual ETC mechanism for the corresponding
fermion mass generation would not apply.

This research was partially supported by the grants DE-FG02-92ER-4074 (T.A.),
NSF-PHY-00-98527 (R.S.).

\vfill
\eject


\begin{thebibliography}{99}
\bibitem{data}{Current data implies that, for $g_{2R} \simeq g_{2L}$,
$m_{W_R} \gsim 800$ GeV, with a similar lower bound on an $m_{Z^\prime}$
\cite{pdg}.}
\bibitem{vals}{A vectorial SU($N$) theory with $N_f$ massless fermions in the
fundamental representation is expected to exist in a confining phase with
S$\chi$SB if $N_f < N_{f,cond.}$, where $N_{f,cond.} \simeq
(2/5)N(50N^2-33)/(5N^2-3)$ and in a nonabelian Coulomb phase if
$N_{f,cond.} < N_f < 11N/2$.  For $N = 2$, we have $N_{f,cond.} \simeq 8$.}
\bibitem{gap}{In the approximation of single-gauge-boson exchange, the
critical coupling for condensation $R_1 \times R_2 \to R_{cond.}$ is given by
$\frac{3\alpha}{2\pi}\Delta C_2 = 1$, where $\Delta C_2 =
[C_2(R_1)+C_2(R_2)-C_2(R_{cond.})]$ and $C_2(R)$ is the quadratic Casimir.}
\bibitem{rh}{We write SM-singlet fields as right-handed.}
\bibitem{GMAC} {This problem is currently under study. The analysis is more
challenging than perturbative vacuum alignment \cite{valign} since all the
relevant couplings are strong.}
\bibitem{lcon}{In Ref. \cite{nt}, which did not use a gauged $B-L$ symmetry, we
employed a different convention, assigning $L=1$ to $\psi_{ij,R}$ so that these
Dirac mass terms conserve $L$ and the $\Delta L = 2$ violation was manifest in
induced $\psi^T_{1i,R} C \psi_{1j,R}$ operators as well as left-handed Majorana
bilinears.}
\end{thebibliography}
\end{document}